\documentclass[structabstract]{aa} 

\usepackage{graphicx}
\usepackage{amsmath}
\usepackage{txfonts}
\usepackage{natbib}
\usepackage{xspace}
\usepackage{textcomp}
\usepackage{dsfont}
\usepackage{color}

\makeatletter
\newlength{\figwidth}
\if@referee
\newcommand{\hl}[1]{\textbf{#1}}
\else
\newcommand{\hl}[1]{#1}
\fi
\makeatother

\newcommand{\bo}{\ensuremath{\boldsymbol{B}_0}}
\newcommand{\const}{\ensuremath{\text{const}}}
\newcommand{\abs}[1]{\ensuremath{\left\lvert #1\right\rvert}}

\newcommand{\eps}{\ensuremath{\varepsilon}}

\newcommand{\be}{\begin{equation}}
\newcommand{\ee}{\end{equation}}
\newcommand{\bs}{\begin{subequations}}
\newcommand{\es}{\end{subequations}}

\newcommand{\R}{\ensuremath{\mathds{R}}}

\newcommand{\pa}{\ensuremath{_\parallel}}
\newcommand{\se}{\ensuremath{_\perp}}

\newcommand{\De}{\ensuremath{\varDelta}}
\newcommand{\Om}{\ensuremath{\varOmega}}

\newcommand{\uint}{\ensuremath{\int_{-\infty}^\infty}}

\newcommand{\f}[1]{\ensuremath{\boldsymbol{#1}}}
\newcommand{\m}[1]{\ensuremath{\left\langle #1\right\rangle}}

\newcommand{\pd}[2][]{\ensuremath{\frac{\partial #1}{\partial #2}}}
\newcommand{\dd}[2][]{\ensuremath{\frac{\mathrm{d} #1}{\mathrm{d} #2}}}
\newcommand{\df}{\ensuremath{\mathrm{d}}}

\newcommand{\etal}{ \emph{et\,al.}}

\begin{document}
\title{Monte-Carlo simulations of intensity profiles\\for energetic particle propagation}
\author{R.\,C. Tautz\inst{1}, J. Bolte\inst{1}, and A. Shalchi\inst{2}}
\institute{Zentrum f\"ur Astronomie und Astrophysik, Technische Universit\"at Berlin, Hardenbergstra\ss{}e 36, D-10623 Berlin, Germany.\\\email{robert.c.tautz@gmail.com}
\and Department of Physics and Astronomy, University of Manitoba, Winnipeg, Manitoba R3T 2N2, Canada. \email{andreasm4@yahoo.com}}
\date{Received August 26, 2015; accepted November 24, 2015}

\abstract
{}
{Numerical test-particle simulations are a reliable and frequently used tool to test analytical transport theories and to predict mean-free paths. The comparison between solutions of the diffusion equation and the particle flux is used to critically judge the applicability of diffusion to the stochastic transport of energetic particles in magnetized turbulence.}
{A Monte-Carlo simulation code is extended to allow for the generation of intensity profiles as well as anisotropy-time profiles. Due to the relatively low number density of computational particles, a kernel function has to be used to describe the spatial extent of each particle.}
{The obtained intensity profiles are interpreted as solutions of the diffusion equation by inserting the diffusion coefficients that have been directly determined from the mean-square displacements. The comparison shows that the time dependence of the diffusion coefficients needs to be considered, in particular the initial ballistic phase and the often sub-diffusive perpendicular coefficient.}
{It is argued that the perpendicular component of the distribution function is essential if agreement between the diffusion solution and the simulated flux is to be obtained. In addition, time-dependent diffusion can provide a better description than the classic diffusion equation only after the initial ballistic phase.}

\keywords{plasmas --- diffusion --- methods: numerical --- turbulence}
\authorrunning{Tautz\etal}
\titlerunning{Monte-Carlo simulations of intensity profiles}
\maketitle

\section{Introduction}

The interplanetary medium has been, and continues to be, one of the most interesting plasmas to be studied both experimentally and theoretically. The main reasons are that the heliosphere is accessible to \emph{in-situ} measurements by spacecrafts, and that Earth-bound observations can be done at relatively low cost. A prominent example for such investigations are solar energetic particles \citep[e.\,g.,][]{lar13:int}, which are not only interesting for basic research but which also affect us economically via the so-called space weather \citep[see][for an introduction]{sch05:wea}.

In the past 50~years, considerable efforts have been put into the determination and prediction of energetic particle propagation in the turbulent solar wind. Accordingly, intensity profiles have been fitted to many models on a variety of refinement levels. These include transport equations---such as Parker's \citeyearpar{par65:pas} transport equation or the \citet{roe69:int} equation---that are based on diffusion and convection \citep[e.\,g.,][]{gom81:fla,pal82:con,dro09:tes,art11:dif}. In essence, a diffusive behavior of the energetic particles has been assumed so that the diffusion coefficients are free fit parameters.

Since the 1990s, numerical test-particle simulations have been used as an alternative method to determine diffusion coefficients \citep[e.\,g.,][and references therein]{mic96:alf,gia99:sim,lai13:cro,tau13:num}. Because kinetic plasma theory has shown that the resulting diffusion coefficients depend sensitively on electromagnetic turbulence \citep[see][for an introduction]{rs:rays,sha09:nli}, the diffusion coefficients can then be used to extract information about the nature of the turbulent interplanetary medium.

However, a careful inspection especially of early analytical transport theories such as quasi-linear theory \citep{jok66:qlt} reveals that, depending on the level of approximation, invalid results are found \citep[e.\,g.,][]{tau06:sta}. Exceptions are the second-order quasi-linear theory \citep{sha05:soq,tau08:soq} and the unified non-linear theory \citep{sha10:uni,sha15:per}, which show remarkable agreement with simulations. Every additional turbulence effect that is included in an analytical description has the potential to significantly alter the results.

Accordingly, the importance of a detailed understanding of the solar wind turbulence cannot be overstated, and many models have tried to incorporate (i) the anisotropy between the directions parallel and perpendicular to the mean magnetic field \citep[e.\,g.,][]{bie96:two,mat90:mal}; (ii) dynamical effects \citep{sau99:lft,bel71:alf}; (iii) the effects of intermittency \citep{osm12:int,wan14:int}; and (iv) the dissipation of magnetic structures \citep{col68:sol,hah13:alf,sah09:dis}.

In recent decades, however, it was realized that the transport of energetic charged particles in a turbulent magnetized medium is not always diffusive. This so-called anomalous diffusion refers to a time-dependent diffusion coefficient in the form $\kappa\propto t^\gamma$ and is characterized as sub-diffusive ($\gamma<1$) and super-diffusive ($\gamma>1$). Indeed, sub-diffusive and sometimes super-diffusive behavior has been found on many occasions \citep[e.\,g.,][]{qin02b:sim,zim06:dif,sha07:flr,tau10:wav,tau10:sub}. New theoretical models are able to take these effects into account \citep[e.\,g.,][]{zim05:ano,pom07:ano,sha11:sub} but that leads to considerably more complicated expressions for the diffusion coefficients. Matters only become worse if others than the most basic turbulence models are to be considered. Explanations of this so-called anomalous transport in terms of the physical origin range from particles back-tracing their magnetic field lines in the case of magnetostatic turbulence---which is indeed the foundation of the compound diffusion model \citep[e.\,g.,][]{kot00:com,web06:com,tau08:sem}---to L\'evy flights in the case of intermittend turbulent structures \citep[e.\,g.,][]{zim05:ano}.

In this article, the two approaches of determining diffusion coefficients---via the mean-square displacement of test particles and via the fit of intensity profiles---will be combined for the first time. Using the Monte-Carlo simulation code \textsc{Padian} that was developed by one of us \citep{tau10:pad}, diffusion coefficients can be determined from the mean-square displacement of test particles moving in magnetic turbulence. At the same time, intensity profiles or, equivalently, the particle flux can be recorded at a given distance from the particle source. The so obtained (directional-averaged) intensity profiles are then compared to diffusion models with the diffusion coefficients: (i) either as fit parameters or (ii) as obtained directly from the simulations. In essence, the approach allows for the simulation of time profiles in a situation where the exact diffusion coefficients are already known.

In addition, the pitch-angle dependence of the incoming particles can also be resolved. Such may be used mainly for two reasons: (i) to trace the isotropization and so to allow a comparison with theoretical expectations and analytical calculations \citep[e.\,g.,][]{dro09:tes}; (ii) to compare simulation results directly to sector data, e.\,g., on the EPAC instrument \citep{kep95:uly} on board the Ulysses spacecraft \citep{kep92:uly} and the Low-Energy Charged Particle instrument on board Voyager~1 \citep[see, e.\,g.,][for a recent application]{dec12:mer}.

The paper is organized as follows. In Sec.~\ref{padian}, the numerical Monte-Carlo code is described that will be used to evaluate the trajectories of test particles under the influence of a turbulent magnetic field. The numerical techniques required to record the flux at a given distance from the particle source are described in Sec.~\ref{aniso}. In Sec.~\ref{timedep}, results will be shown and the possible agreement with the solution of a generalized diffusion equation will be discussed. In Sec.~\ref{summ}, a summary of the results and a discussion of further applications conclude this paper.

\section{\textsc{Padian} test-particle code}\label{padian}

For the numerical simulations, the \textsc{Padian} Monte-Carlo code \citep{tau10:pad} is used to compute the parallel and perpendicular diffusion coefficient and the time profiles of energetic particles. For the turbulent electromagnetic fields, the Fourier superposition model is used \citep{bat82:tur,gia99:sim,tau13:num}, which allows the use of a power spectrum that is based on observations and/or turbulence theory. The geometry of the turbulent fields, which are superimposed on a homogeneous mean magnetic field \bo, is assumed to be isotropic so that no preferred direction exists for the wave vectors of the Fourier components. The corresponding generation of turbulent magnetic fields proceeds as \citep{tau13:num}
\be\label{eq:dB}
\delta\f B(\f r,t)=\sum_{n=1}^N\f e'_\perp A(k_n)\cos\left[k_nz'+\beta_n-\omega(k_n)t\right],
\ee
where $\beta$ is a random phase angle and where the primed coordinates are obtained from a rotation with random angles. This allows one to include also more realistic turbulence geometries based on the observed anisotropies in the solar wind \citep[e.\,g.,][]{bie96:two,mat90:mal,rau13:mal}.

The wavenumbers $k_n$ are distributed logarithmically in the interval $k_{\text{min}} \leqslant k_n \leqslant k_{\text{max}}$. For the minimum and maximum wavenumbers included in the turbulence generator, the following considerations apply: (i) the \emph{resonance condition} states that there has to be a parallel wavenumber $k\pa$ so that $R_{\text L}\mu k\pa=1$ with $\mu=\cos\angle(\f v,\bo)$ and where $R_{\text L}$ denotes the particle's Larmor radius. Thus, scattering predominantly occurs when a particle can interact with a wave mode over a full gyration cycle; (ii) the \emph{scaling condition} requires that $R_{\text L}\Om_{\text{rel}}t<L_{\text{max}}$, where $\Om_{\text{rel}}=qB/(\gamma mc)$ is the relativistic gyrofrequency and where $L_{\text{max}}\propto1/k_{\text{min}}$ is the maximum extension of the system, which is given by the lowest wavenumber (for which one has $k_{\text{min}}=2\pi/\lambda_{\text{max}}$, thereby proving the argument). In practice, the second condition determines the minimum 
wavenumber while the first one determines also the maximum wavenumber. Here, values are chosen as $k_{\text{min}}\ell_0=10^{-5}$ and $k_{\text{max}}\ell_0=10^3$, where $\ell_0$ is the turbulence bend-over scale. The sum in Eq.~\eqref{eq:dB} extends over $N=512$ wave modes, which is sufficient \citep{tau13:num} and yet saves computation time. Furthermore, the maximum simulation time is determined as $vt_{\text{max}}/\ell_0=10^2$, which has the advantage that, by using the so normalized simulation time, the behavior of the scattering parameters becomes mostly independent of the particle energy.

\begin{figure}[t]
\centering
\includegraphics[bb=160 262 430 578,clip,width=\figwidth]{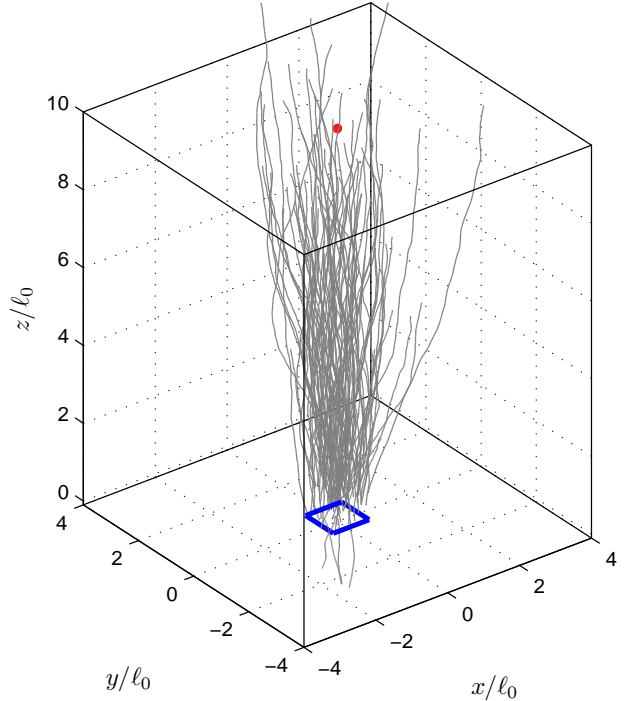}
\caption{(Color online) Sample trajectories starting from the rectangular blue region with an initial pitch-angle cosine of $\mu_0=1$, hence moving in the positive $z$ direction. The target region is shown as the small red dot a $[x,y,z]=[0,0,10]$.
\vspace{1em}}
\label{ab:trajes}
\end{figure}

The polarization vector obeys the condition $\f e'_\perp\cdot\f e'_z=0$, where the primed coordinates are determined via a rotation matrix with random angles so that $\f k$ has a random direction for each wave mode. Alternatively, the inclusion of plasma waves is in principle possible through a dispersion relation $\omega(k)$; here, however, magnetostatic turbulence is used with $\omega\equiv0$.

\begin{figure}[t]
\centering
\includegraphics[width=\figwidth]{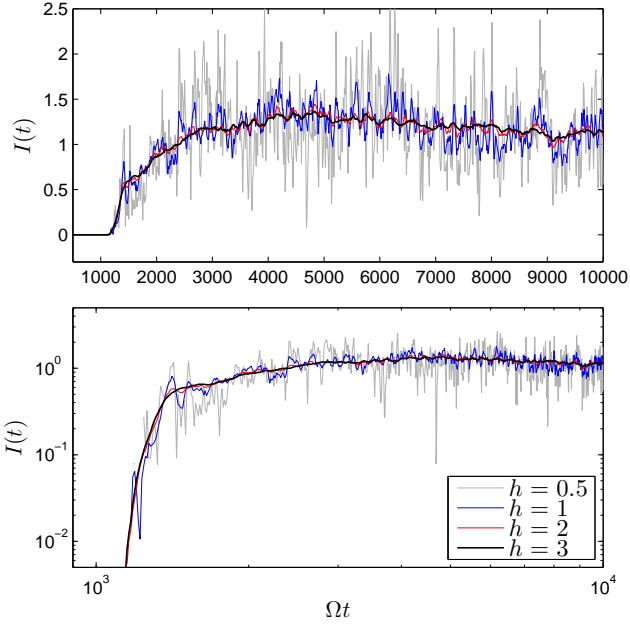}
\caption{(Color online) Intensity profile for particles with rigidities $R=10^{-1}$ and a distance between the particle source and the detector of $L=100\ell_0$ in linear (upper panel) and logarithmic (lower panel) units. In the SPH kernel according to Eq.~\eqref{eq:sph}, the base smoothing length is varied between $h=0.5\ell_0$ and $3\ell_0$ (see legend). An additional influence on the smoothness of the curve can be obtained through the detector size. As illustrated, the smoothing length modifies the fluctuation of the intensity profile but otherwise leaves the underlying structure unchanged.
\vspace{1em}}
\label{ab:smoothing}
\end{figure}

The amplitude function, $A(k_n)\propto\sqrt{G(k_n)\,\De k_n}$, is proportional to the square root of the turbulence power spectrum, $G(k)$, for which a kappa-type function is used \citep{sha09:flr} as
\be\label{eq:spect}
G(k)\propto\frac{\abs{\ell_0k}^q}{\left[1+(\ell_0k)^2\right]^{(s+q)/2}},
\ee
with $q=0$ for simplicity and to ensure the comparability with earlier results.\footnote{Note that, to ensure that magnetic field lines behave diffusively, a positive $q>1$ would be required \citep{wei10:flr}.} The turbulence bend-over scale, $\ell_0\approx0.03$\,au, reflects the transition from the energy range $G(k)\propto k^q$ to the Kolmogorov-type inertial range, where $G(k)\propto k^{-s}$ with $s=5/3$ \citep{kol41:tur,bru05:sol}. A normalization factor relates the integral over the spectrum to the mean turbulence strength, but is omitted here because Eq.~\eqref{eq:dB} has been designed so that the turbulent field has the correct strength. Note that there are no actual boundary conditions but, instead, care should be taken that particles travel no further than the maximum distance $L_{\text{max}}\propto k_{\text{min}}^{-1}$, beyond which the turbulent magnetic field pattern would repeat itself.

From the integration of the Newton-Lorentz equation and by averaging over an ensemble of particles, the diffusion coefficients can be calculated by determining the mean square displacement as \bs\label{eq:kappa}
\begin{align}
\kappa\pa&=\dd t\m{\left(\De z\right)^2}\approx\frac{1}{2t}\m{\left(\De x_i\right)^2}\\
\kappa\se&=\frac{1}{2}\dd t\m{\left(\De x\right)^2+\left(\De y\right)^2}\approx\frac{1}{4t}\m{\left(\De x\right)^2+\left(\De y\right)^2}.
\end{align}
\es
The scattering mean-free path in the direction parallel to the background magnetic field can then be obtained as $\lambda\pa=3\kappa\pa/v\approx\langle(\De z)^2\rangle/(2vt)$ for large $t$.

\begin{figure}[t]
\centering
\includegraphics[width=\figwidth]{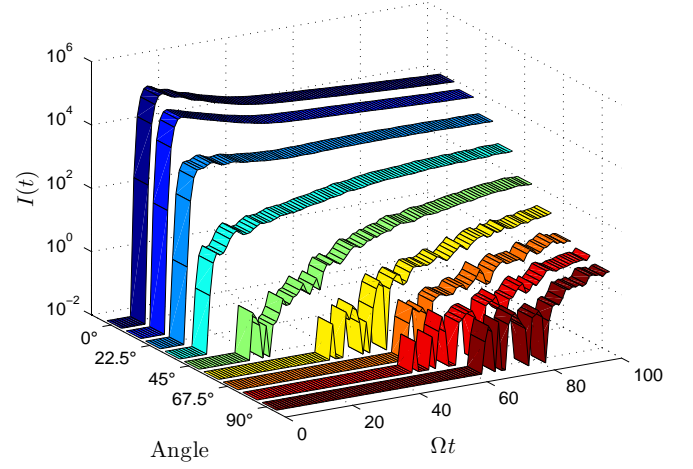}
\caption{(Color online) Intensity-time profiles for particles with rigidities $R=1$ and a distance between the particle source and the detector of $L=10\ell_0$. The effect is shown of a detector position being either along the same field line with the particle source (angle of $0^\circ$) up to a detector perpendicular to the source field line (angle of $90^\circ$).
\vspace{1em}}
\label{ab:angles}
\end{figure}

\section{Intensity and anisotropy-time profiles}\label{aniso}

In Fig.~\ref{ab:trajes}, a representative sample of particle trajectories is illustrated for typical particle energies ranging from $-4\leqslant\log R\leqslant-2$ with $R=\gamma v/(\Om\ell_0)$ a normalized rigidity variable that is per default used in the \textsc{Padian} code \citep{tau10:pad}. All---typically $10^5$ up to $10^7$---test particles are injected instantaneously at time $t=0$ with random initial positions inside a cubic region with an edge length of $0.1\ell_0$. Accordingly, all length scales are normalized to the turbulence bend-over scale, $\ell_0$, and two dimensionless times are introduced as $\Om t$ and $vt/\ell_0$. The relative strength of the isotropic turbulent magnetic fields are chosen as $(\delta B/B_0)^2=0.1$ with $B_0$ the background magnetic field strength. Note that, in the solar wind, values of up to $\delta B/B_0=1$ have been observed. Therefore, a relative turbulence strength of $0.1$ is an acceptable compromise between the particles still moving predominantly along the mean magnetic field (which is helpful in order to obtain good statistics) and a sufficiently strong stochastic motion. In the range $0<\delta B/B_0\leqslant1$, it is to be expected that the results do not vary apart from some rescaling.

The anisotropy-time profile is then measured at a virtual spherical detector (red dot in Fig.~\ref{ab:trajes}), the position of which can be varied. In what follows, the numerical technique will be described that allows the evaluation of time profiles, which will then be compared to solutions of the (generalized) diffusion equation in Sec.~\ref{timedep}.

\subsection{SPH smoothing}

Within the test-particle simulations, a time profile can, in principle, be obtained by counting the rate of particles crossing the surface of a particular region of space (called the ``detector''). However, due to limitations of computational power, the particle flux at any reasonable distance between the (common) particle origin and the detector would be too low even for $10^7$ particles because perpendicular scattering, albeit weak, significantly reduces the number of particles arriving at the detector. A naive solution would be to increase the detector size drastically, in which case however much of the fine structure would be either lost or smeared out. In addition, the onset time would be significantly reduced if the detector size is too large.

\begin{figure}[t]
\centering
\includegraphics[bb=105 270 485 571,width=\figwidth]{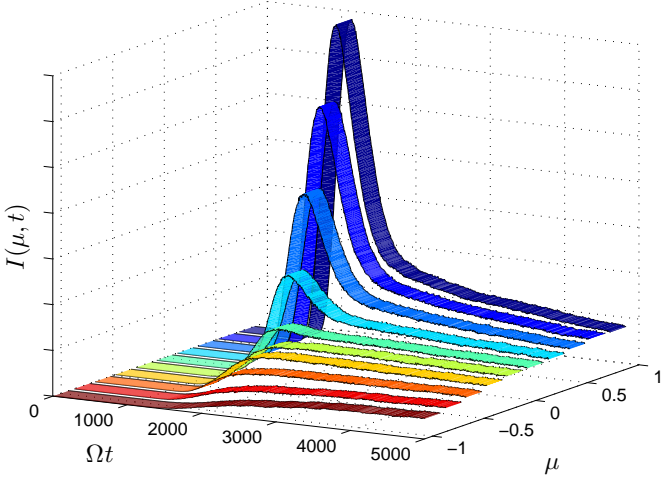}
\caption{(Color online) Pitch-angle dependent intensity profile for particles with a normalized rigidity $R=10^{-2}$ in arbitrary but linear units. Shown is the flux of particles arriving at the detector with $\mu$ ranging from $+1$ (i.\,e., coming from the direction of the source) to $-1$ (coming from the far side of the source).
\vspace{1em}}
\label{ab:prof_pita}
\end{figure}

Alternatively, the following will be inspired by the smoothed particle hydrodynamics (SPH) approach \citep[e.\,g.,][]{vau08:sph,spr10:sph}. The basic idea is that all computational particles are no longer treated as pointlike physical particles. Instead, each computational particle represents an ensemble of real particles that is distributed smoothly over a defined region. A typical density distribution (the so-called SPH kernel) is given by
\be\label{eq:sph}
f_{\text{SPH}}(\f r)=
\left\{
\begin{tabular}{ll}
$\left(4-6q^2+3q^3\right)/(6h)$,	& $q\leqslant1$\\
$\left(2-q\right)^3/(6h)$,		& $q\leqslant2$
\end{tabular}
\right.,
\ee
and zero otherwise, where $q=\lvert\f r-\f r_{\text{target}}\rvert/h$ with $h$ the smoothing length. Such a distribution is similar in shape to a Gaussian distribution but, at the same time, avoids the problem of a finite density even at very large distances.

To avoid the problem of continuously decreasing fluxes for increasing distances, the detector---or target region---generally scales with the distance between target and particle origin. As shown in Fig.~\ref{ab:smoothing}, a larger smoothing length is able to filter the high-frequency fluctuations effectively, while leaving the overall shape of the intensity profile unchanged.

In addition, it has to be noted that, for a relatively small detector size such as that used in Fig.~\ref{ab:smoothing}, almost no signal would be visible \emph{without} the SPH ansatz. Therefore, it is precisely this method that allows for the generation of intensity profiles at all, even though some of the curves look rather noisy.

\subsection{Sample results}

For the evaluation of a sample time profile, a distance of $L=100\ell_0$ (which roughly corresponds to $\approx45\times10^7$\,km) has been chosen between the particle source and the detector.
\hl{The target size is $d=0.1\ell_0$,}
thus representing a compromise between the ratio of distance to target size being at least 1000~times smaller than $L$ but, at the same time, being able to accumulate a sufficiently large number of particles. The SPH smoothing length is chosen so that each numerical particle represents a particle ensemble that is distributed over 10--60~times the detector size according to Eq.~\eqref{eq:sph}.

\begin{figure}[t]
\centering
\includegraphics[bb=105 211 485 630,width=\figwidth]{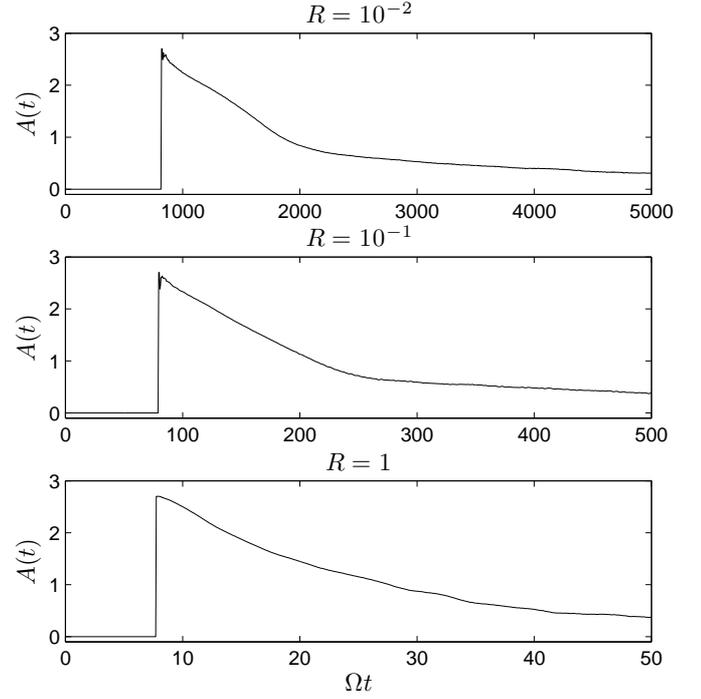}
\caption{Anisotropy as obtained by taking the first and zeroth moment of the SPH-smoothed profiles according to Eq.~\eqref{eq:ani}. Note that, for particles with different rigidities (see panel titles), the time scales are dramatically different.
\vspace{1em}}
\label{ab:prof_pitb}
\end{figure}

In Fig.~\ref{ab:smoothing}, a typical intensity profiles is shown and the effect of the SPH smoothing is illustrated. This technique is necessary to reduce the noise that arises from the (relatively) low number of simulated particles. The actual total counts (fluence) can be as low as $<10$ for a low rigidity and $<100$ for the highest rigidities considered. In general, it can be found that particles with higher energies arrive earlier, even though they experience significantly stronger scattering.

If the detector position is not along the same field line as the particle origin---which in reality should be rather the rule than the exception---the count rates are strongly reduced and the shape of the intensity profile is changed, too. In Fig.~\ref{ab:angles}, this effect is illustrated for detector positions varying between $0^\circ$ (i.\,e., being along the field line of the particle source) or $90^\circ$ (i.\,e., being perpendicular to it). For such particles, it takes longer to begin reaching the detector \citep[cf.][]{qin11:con}. Even though for the shown example a small distance between the particle source and the detector and a relatively high particle energy are chosen, $10^7$ test particles were required in order to resolve the intensity profiles at large angles.

\subsection{Pitch-angle dependency and anisotropy}

Normally, the limited number of simulated particles does not allow to resolve the pitch-angle dependence of the incoming particles, at least not in the form of a two-dimensional intensity profile. Therefore, a tool such as SPH smoothing is mandatory as it allows to assign a broad density distribution to an otherwise point-like particle.

In the literature, a commonly used measure is the time-dependent anisotropy \citep[e.\,g.,][]{sch09:atp}, which is defined via the first moment in the pitch-angle cosine as
\be\label{eq:ani}
A(t)=3\int_{-1}^1\df\mu\;\mu I(\mu,t)\left/\int_{-1}^1\df\mu\;I(\mu,t)\right..
\ee
Here, $I(\mu,t)$ is the pitch-angle-dependent distribution function, which, for each $\mu$ value, is obtained with the help of SPH smoothing. The so defined anisotropy function provides a powerful diagnostics tool that enables one to probe heliospheric particle scattering parameters such as diffusion coefficients and mean-free paths as well as drift coefficients \citep[e.\,g.,][]{dro05:pro,sch09:atp}. In addition, the anisotropy can be related to a spatial gradient of the particle density \citep{sch89:cr2,sha09:hil}.

The pitch-angle dependent time profile together with the asymmetry defined in Eq.~\eqref{eq:ani} are shown in Figs.~\ref{ab:prof_pita} and \ref{ab:prof_pitb}, respectively. The result confirms the obvious expectations that, during the initial burst, particles predominantly arrive from the direction of the source. For later times the asymmetry remains positive, which underlines that there are always more particles moving outward.

By repeating the simulation for different particle rigidities varying between $R=10^{-2}$ and $R=1$ and by noting the results shown in Fig.~\ref{ab:prof_pitb} it can be seen that

\begin{itemize}
\item low-energy particles arrive predominantly from the direction of the source (i.\,e., $A>0$) for a long time;
\item at short distances, high-energy particles have a higher probability of being back scattered behind the detector (i.\,e., $A\to0$);
\item in the limit of large times, particles tend to fill space homogeneously and move isotropically.
\end{itemize}

\section{Time-dependent diffusion}\label{timedep}

\begin{figure}[t]
\centering
\includegraphics[width=\figwidth]{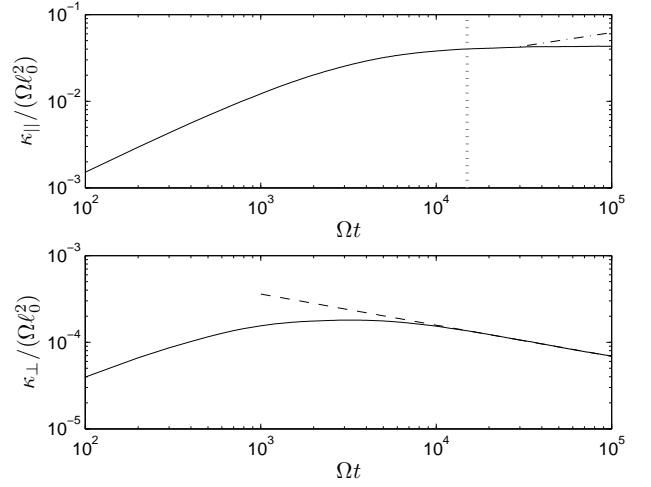}
\caption{Example for the parallel (upper panel) and perpendicular (lower panel) diffusion coefficients. During the ballistic phase, the diffusion coefficients grow almost linearly in time, whereas for late times the parallel diffusion coefficient becomes almost constant. For the parallel diffusion coefficient, the transition is marked for later use by the vertical dashed line. In addition, the dot-dashed line illustrates a late super-diffusive phase with $\kappa\pa\propto t^{0.31}$, which has been observed for Alfv\'enic turbulence due to stochastic acceleration of the particles \citep[cf.][]{tau10:wav}.
\hl{Here, in contrast, a magnetostatic model is used, for which the solid line is found.}
The perpendicular diffusion coefficients, in contrast, become slightly sub-diffusive in agreement with earlier results. For later use, the late-time behavior is fitted to a power law. Note that the slope sensitively depends on the properties of the magnetic turbulence \citep[see][]{tau10:sub} and the particle rigidity (here, $R=10^{-2}$).
\vspace{1em}}
\label{ab:tdif}
\end{figure}

In this section, the interpretations of the time profiles will be discussed that are obtained from the virtual detector within the \textsc{Padian} simulation. As will be shown, the comparison of a recorded intensity profile to a solution of the diffusion equation can be impeded mainly by two factors: (i) the early phase cannot be correctly described by a diffusion approach; and (ii) the diffusion coefficients of cosmic-ray scattering are often time-dependent also for \emph{late} times. The first factor has been widely acknowledged and, accordingly, other theoretical descriptions are in use \citep[e.\,g.,][]{lai13:cro,eff14:te1}.

The second factor is illustrated in Fig.~\ref{ab:tdif} and shows first the initial ballistic phase with free particle motion so that $\kappa\propto t$. After that, either the typically diffusive or sometimes super-diffusive (for parallel transport) or mostly sub-diffusive (for perpendicular transport) phases are reached. If turbulent electric fields were to be included that arise, for instance, in Alfv\'enic turbulence \citep{lee75:mhd,ach81:fer,mic96:alf}, a super-diffusive behavior has been found \citep{tau10:wav}, which can be explained with the stochastic acceleration of particles as a result of momentum diffusion \citep[e.\,g.,][and references therein]{tau13:sto}.

In what follows, the three-dimensional diffusion equation with time-dependent coefficients (see Appendix~\ref{appA}),
\be\label{eq:deq}
\pd[f_{\text{Diff}}]t=\left[\f\kappa(t)\cdot\nabla\right]\cdot\nabla f_{\text{Diff}},
\ee
will be employed, which is the simplest extension of the usual diffusion equation. In the diffusion tensor, the off-diagonal drift coefficients \citep[see, e.\,g.,][]{tau12:kpa} will be neglected, as these are important only for non-turbulent and curved background magnetic fields (see the discussion in Appendix~\ref{appA}). Therefore, the form $\f\kappa=\text{diag}(\kappa\se,\kappa\se,\kappa\pa)$ is employed, with all coefficients being functions of time.

Together with the initial condition $f_{\text{Diff}}(x,y,z,t=0)=\delta^3(x,y,z)$, the solution to Eq.~\eqref{eq:deq} provides the time profile function $f_{\text{Diff}}(x,y,z,t)$. Note that, due to the finite size of the source region, the point-source solution in Eq.~\eqref{eq:deq} would have to be convolved with the shape of the source region. However, it turns out that the resulting corrections are negligible due to the small extension of the source region. The relevant fits that will be shown are: (i) solution of the three-dimensional diffusion equation with the time-dependent diffusion coefficients, $\kappa(t)$, obtained from the simulations via the mean-square displacements; (ii) solution of the three-dimensional diffusion equation with constant diffusion coefficients, $\kappa(t_{\text{max}})$ i.\,e., the values at the end of the simulations.

\subsection{Detector at small distances}

\begin{figure}[t]
\centering
\includegraphics[bb=103 175 487 665,width=\figwidth]{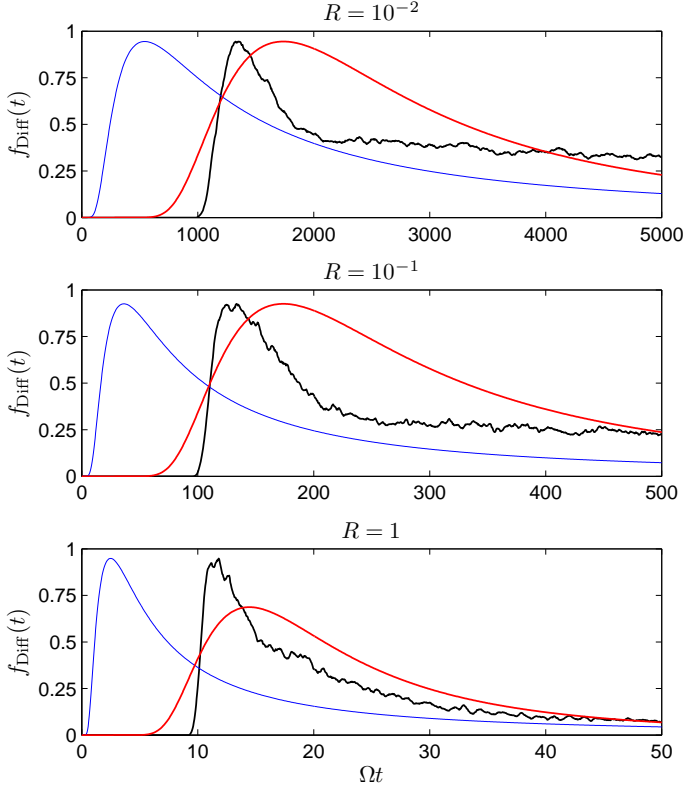}
\caption{(Color online) Time profile (intensity profile) in arbitrary linear units for particles with variable energies (see subplot titles). The distance between the particle origin and the virtual detector is $L=10\ell_0$. The various curves are as follows: The black lines shows the profile as obtained from the simulations using the methods described in Sec.~\ref{aniso}. The red lines show the three-dimensional diffusion solution (with the normalization factor as an open parameter) from Eq.~\eqref{eq:diff3D} using the diffusion coefficients as obtained from the simulations via the mean-square displacements. In addition, the result from Eq.~\eqref{eq:diff3DN} is shown for \emph{constant} diffusion coefficients taken from the simulations (blue lines), i.\,e., $\kappa_i(t=t_{\text{max}})$.
\vspace{1em}}
\label{ab:kappafit1}
\end{figure}

If, for the case of instantaneous injection, the distance between the particle source and the detector is small, the particles will still be in the ballistic phase, where $\kappa\propto t$, when the intensity profile reaches its maximum. The time-dependent diffusion coefficient can therefore be written in the form of a single power law, i.\,e., $\kappa=\xi t^\gamma$. For this case, a solution to Eq.~\eqref{eq:deq} can be obtained analytically (see Appendix~\ref{appA}) as
\be\label{eq:diff3D}
f_{\text{Diff}}(\f r,t)=\prod_i\frac{1}{\sqrt{4\pi\xi_it^{\gamma_i+1}}}\,\exp\!\left[-\frac{\left(1+\gamma_i\right){r_i}^2}{4\xi_it^{\gamma_i+1}}\right]
\ee
for $i\in\{x,y,z\}$. The diffusion coefficients are taken directly from the simulated particle trajectories via the mean-square displacement, see Eq.~\eqref{eq:kappa}, as $\kappa_i=\m{(\De r_i)^2}/(2t)$ with $i\in\{x,y,z\}$ and are then fitted to a power law. In practice, this is done by applying a linear regression to the logarithm of $\kappa$. It should be noted, however, that this fit is not always unique because, even during the initial phase, there can be some variations and irregularities especially in the perpendicular diffusion coefficients. After that, typical values are $-0.05\leqslant\gamma\leqslant-0.25$, showing that, in isotropic turbulence, perpendicular transport is only mildly subdiffusive \citep[cf.][]{tau10:sub}.

For a detector distance $L=10\ell_0$, the resulting function $f_{\text{Diff}}$ from Eq.~\eqref{eq:diff3D} together with the time profile recorded within the simulation are shown in Fig.~\ref{ab:kappafit1}. Both for the overall shape as well as the peak intensity, a rather poor agreement can be found, regardless of the particle rigidity. The reason is that, during the ballistic phase, the particle motion \emph{cannot} be described in terms of a (super) diffusive behavior (see Sec.~\ref{disc}).

Nevertheless, it has to be noted that, by neglecting the time dependence of the diffusion coefficients, there is even less agreement with the simulations. This has been checked by using the last value for the diffusion coefficient as obtained from the simulations, i.\,e., $\kappa_i=\kappa_i(t=t_{\text{max}})=\const$. In addition, $\gamma_i=0$ is used in Eq.~\eqref{eq:diff3D} so that $f_{\text{Diff}}$ corresponds to the classic diffusion solution. As shown in Fig.~\ref{ab:kappafit1}, such is clearly not an option.

\subsection{Detector at large distances}

\begin{figure}[t]
\centering
\includegraphics[bb=103 175 487 665,width=\figwidth]{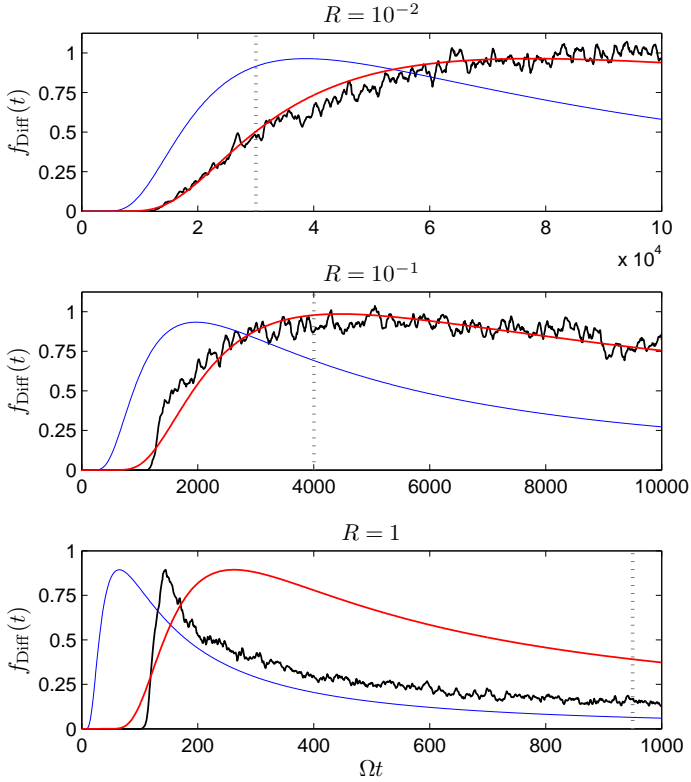}
\caption{(Color online) Same as Fig.~\ref{ab:kappafit1} but for a distance $L=100\ell_0$ between the particle origin and the virtual detector. The vertical dashed lines mark the end of the initial ballistic regime as explained in Fig.~\ref{ab:tdif}.
\vspace{1em}}
\label{ab:kappafit2}
\end{figure}

\begin{figure}[t]
\centering
\includegraphics[bb=103 175 487 665,width=\figwidth]{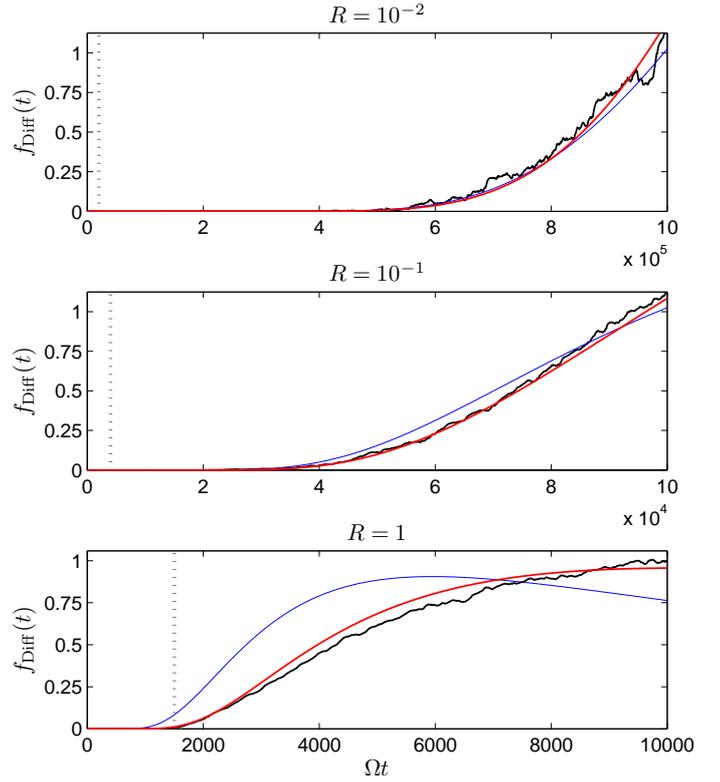}
\caption{(Color online) Same as Fig.~\ref{ab:kappafit2} but for a distance $L=1000\ell_0$ between the particle origin and the virtual detector.
\vspace{1em}}
\label{ab:kappafit3}
\end{figure}

For general time-dependent diffusion coefficients, there seems to be no simple way to solve Eq.~\eqref{eq:deq} analytically (Appendix~\ref{appA}). Therefore, a numerical approach is chosen (Appendix~\ref{appB}). To avoid the numerically challenging solution in three spatial dimensions, the perpendicular diffusion coefficients are assumed to follow a single power law, which will be that at late times. This is supported by the fact that, for the purpose of this investigation, the detector is placed on the same field line as the particle source so that the contribution from perpendicular diffusion does not appear in the exponential function.

In accordance with Eq.~\eqref{eq:diff3D}, the assumed three-dimensional solution function therefore has the form
\begin{align}
f_{\text{Diff}}(x,y,z,t)&=f\pa(z,t)\nonumber\\
\times\prod_i&\frac{1}{\sqrt{4\pi\xi_it^{\gamma_i+1}}}\exp\!\left[-\frac{\left(1+\gamma_i\right){r_i}^2}{4\xi_it^{\gamma_i+1}}\right], \label{eq:diff3DN}
\end{align}
this time for $i\in\{x,y\}$ only. In Eq.~\eqref{eq:diff3DN}, the parallel part, $f\pa(z,t)$ is obtained by numerically integrating the one-dimensional diffusion equation with a time-dependent diffusion coefficient, $\kappa\pa(t)$, as described in Appendix~\ref{appB}.

In Fig.~\ref{ab:kappafit2}, a comparison is shown between the intensity profile obtained within the simulations and the semi-numerical solution of Eq.~\eqref{eq:diff3DN} for a detector distance $L=100\ell_0$. Depending on the particle energy, the end of the ballistic phase is reached sooner---before the peak intensity is reached---or later. It is precisely this point which decides if the solution from Eq.~\eqref{eq:diff3DN} is in agreement with the numerical time profile. In any case, no agreement is found with the solution that uses a constant diffusion coefficient.

In Fig.~\ref{ab:kappafit3}, the extreme case is shown with a detector distance $L=1000\ell_0$ corresponding to $\sim\!30$\,au, which distance is of the order of the heliosphere itself and might therefore be applicable for anomalous cosmic rays. Here, the ballistic phase ends before a noticeable fraction of the particles reaches the detector. Accordingly, the solution in Eq.~\eqref{eq:diff3DN} is able to reproduce the simulated intensity profile with a good accuracy. However, only in the case of the lowest particle energy can there be found agreement with the solution that uses a constant diffusion coefficient.

\subsection{Discussion}\label{disc}

There are three points that are crucial in order to achieve agreement between intensity profile obtained directly from the simulated particle trajectories and the solution of the generalized diffusion equation.

First, the coefficients of perpendicular diffusion are important even though they are by two to four orders of magnitude smaller than the parallel diffusion coefficient. The reason is found in the three-dimensional diffusion solution, Eq.~\eqref{eq:diff3DN}, in which the perpendicular components assure a certain agreement with the simulated intensity profiles. For instance, the shape of the profile is influenced by the product of the diffusion coefficients with the peak intensity being found for $t_{\text{max}}=L^2/(6\kappa\pa)$ at a position along the mean field. (In the one-dimensional case, the factor 6 is replaced by 2.)

Second, the ballistic phase has to be left relatively early---compared to the maximum of the intensity profile---if the intensity profile is to be described by the generalized diffusion approach. As \citet{lai13:cro} noted, early-time particle propagation is dominated by field-line meandering instead of diffusion. The same is also seen for thermal conduction on small scales \citep{huy15:con}. Such could be decided using a quantitative measure for Gaussian diffusion like the kurtosis of the field-line displacement \citep{zim00:kub}. Because for early times---referring to Fig.~\ref{ab:tdif}---the propagation proceeds undisturbed along the magnetic field lines, it is well known that diffusion is merely a time asymptotic model. This explains the poor agreement found in Fig.~\ref{ab:kappafit1} even for time-dependent diffusion coefficients, because the particle motion is simply not diffusive. To our knowledge, there is yet no all-encompassing theoretical description that combines all phases of particle transport.

Third, when the ballistic phase is left, the agreement nevertheless relies on a correct incorporation of the time-dependent diffusion coefficients, as clearly shown in Figs.~\ref{ab:kappafit2} and \ref{ab:kappafit3}. The assumption of constant diffusion coefficients was justified only in those cases, where the first particles arrived at the detector only after the transport had well reached the diffusive regime. One notes from Figs.~\ref{ab:kappafit1} and \ref{ab:kappafit2} that the solutions with constant diffusion coefficients might agree better with the observed profiles if they are shifted to later times. Curiously, the delay time corresponds roughly to $L/(\ell_0R)$, which is the time required for a direct flight from the source region to the detector. Mathematically, however, such would not be a solution of the diffusion equation.

\begin{table}[b]
\centering
\begin{tabular}{llllll}\hline\hline
$L/\ell_0$	& $R$ 		& ${\kappa\pa}^{\text{sim}}$	& $\sigma$	& ${\kappa\pa}^{\text{fit}}$	& $\sigma$\\\hline
$10$			& $10^{-2}$ 	& $0.031$					& $0.005$		& $0.879$					& $0.016$\\
$10$			& $10^{-1}$	& $0.459$					& $0.055$		& $10.43$					& $0.243$\\
$10$			& $1$		& $6.697$   				& $0.834$		& $129.9$					& $3.988$\\
$100$		& $10^{-2}$ 	& $0.043$  				& $0.009$		& $0.023$					& $6.64\times10^{-5}$\\
$100$		& $10^{-1}$	& $0.844$					& $0.160$		& $0.359$					& $1.60\times10^{-3}$\\
$100$		& $1$		& $25.75$   				& $4.478$		& $8.714$					& $0.135$\\
$1000$		& $10^{-2}$ 	& $0.043$  				& $0.009$		& $3.56\times10^{-4}$		& $6.94\times10^{-7}$\\
$1000$		& $10^{-1}$	& $0.871$					& $0.196$		& $6.35\times10^{-3}$		& $1.42\times10^{-5}$\\
$1000$		& $1$		& $28.29$   				& $5.612$		& $0.147$					& $5.79\times10^{-4}$\\
\hline\hline
\end{tabular}
\caption{Comparison between the values obtained for the parallel diffusion coefficients (normalized to $\Om^2\ell_0$) obtained directly from the test-particle trajectories via the mean-square displacements, (${\kappa\pa}^{\text{sim}}$) in Eq.~\eqref{eq:kappa}, and from the fit to a three-dimensional diffusion profile with constant diffusion coefficients, (${\kappa\pa}^{\text{fit}}$), together with their respective estimated mean errors. Note that the use of the SPH smoothing introduces an additional error (not shown here), which however is small compared to the deviation between ${\kappa\pa}^{\text{sim}}$ and ${\kappa\pa}^{\text{fit}}$ and thus does not falsify the analysis.}
\label{ta:comp}
\end{table}

If one were to take the diffusion coefficients as unknown---but time-independent---the time profiles obtained from the simulations can, in some cases, be fitted to a one-dimensional diffusion solution with a constant diffusion coefficient as a fit parameter. The results are shown in Table~\ref{ta:comp} and illustrate that the fitted diffusion coefficients show no agreement with those obtained from the simulations. This underlines again that: (i) a one-dimensional solution is not sufficient despite the significantly smaller perpendicular diffusion coefficients; and (ii) the assumption of a constant diffusion coefficients cannot be justified and introduces significant deviations from the true values. Accordingly, care has to be taken about the correct treatment of the time-dependent diffusion.

\section{Summary and conclusion}\label{summ}

The dilute, magnetized, and turbulent plasma of the heliosphere offers the unique opportunity to investigate fundamental problems such as the transport of energetic particles in two ways. First, it is accessible to in-situ observations of both particles and magnetic fields by various spacecrafts. Second, in combination with increasingly sophisticated numerical simulation techniques, it allows for the critical examination of transport theories.

Throughout the last decades, there has been an increasing insight into the necessity for the kinetic approach for the description of heliospheric and astrophysical plasmas. At the same time, simple heuristic approaches remain in use so that efforts needs to be focused both on the proper understanding and application of the underlying microphysics. In this paper, a connection has been established between the diffusion coefficients obtained by using numerical test-particle simulations and those derived from the observation of intensity profiles at virtual detectors in the same simulations. This approach allows to challenge the seemingly well-established method of fitting the solution of transport equations to obtain (constant) diffusion coefficients.

In future work, we will apply the method to the propagation of solar energetic particle events, which requires the incorporation of a curved mean magnetic field in the form of the Parker spiral \citep{abl15:sol}. In the context of test-particle simulations, this has already been implemented \citep{tau11:spi}. The advantage is that this approach requires no assumptions about the radial dependence of the diffusion coefficients \citep[e.\,g.,][]{sak02:pro}, because they are determined by following the particles as they continue to move outward.

The investigation of solar energetic particle transport poses further challenges. As discussed, e.\,g., by \citet{dre12:spr,lar13:int}, solar particle events can be spread over a very wide range so that a signal can be found even on the opposite side of the field line connected to the particle origin \citep{qin11:con}. Explanations range from strong particle transport perpendicular to the mean magnetic field to an extended energetic particle distribution close to the source but a definitive answer is still elusive. Therefore, the method presented here has the potential to complement other methods, for example by solving a transport equations using finite-difference or stochastic schemes \citep{dro10:tra}.

It has not escaped our attention that the methods outlined in this paper are also applicable to other scenarios. These include the propagation of galactic and anomalous cosmic rays in the outer heliosphere and the transport of galactic and extra-galactic cosmic rays in the interstellar medium. There, however, the dynamics of the turbulent magnetic fields might not be fully captured with a Fourier superposition ansatz so that direct numerical simulations of the turbulence will be required \citep[e.\,g.,][]{avi05:mag}. In addition, the simulation setup also allows for the inclusion of moving detectors that cross, for instant, large-scale and transient structures such as shock waves.

To conclude, it has been shown that anomalous transport regimes play an important role for the interpretation of intensity and, potentially, anisotropy-time profiles, if these are used to extract the diffusion coefficients parallel and perpendicular to the mean magnetic field. The ballistic phase, however, must be left if any agreement is to be expected. The results clearly show that, for the early ballistic phase, an extension of the diffusion ansatz is not permitted and yields wrong results. Even in cases when the bulk of the particles behaves already diffusively in the sense that a time-asymptotic, constant diffusion coefficient is found, in particular the perpendicular sub-diffusive transport has to be considered in order to obtain the correct results. The approach constitutes the simplest analytical continuation toward the initial ballistic phase by using the diffusion approach. As shown, such is permitted if the comparison is done well beyond the ballistic phase. Whether other analytical approaches 
such as the telegraph equation \citep[e.\,g.,][]{lit13:tel,eff14:te1} are able to show better agreement with intensity profiles will be the subject of future work.

\begin{acknowledgements}
RCT acknowledges fruitful discussions with Andreas Kopp. AS acknowledges support by the Natural Sciences and Engineering Research Council (NSERC) of Canada.
\end{acknowledgements}

\appendix

\section{Time-dependent diffusion coefficient: analytical approach}\label{appA}

Analytically, the problem of anomalous diffusion \citep[e.\,g.,][]{zim05:ano,pom07:ano,tau10:sub} has been investigated with a variety of novel approaches. In the most general case, the derivation is based on a fractional Fokker-Planck (or diffusion) equation \citep{sok02:fra,sta03:bro}, which has the general form
\be\label{eq:frac}
\pd[^\alpha f(x,t)]{t^\alpha}=\int_0^t\df t'\:\pd{\abs x}\left[\kappa(x,t-t')\,\pd[^{\mu-1}\bigl(f(x,t')\bigr)^\nu]{\abs{x}^{\mu-1}}\right]
\ee
where $\nu,\mu,\alpha\in\R$ and where $\kappa(x,t)=\kappa(t)\abs{x}^{-\vartheta}$ is the generalized diffusion coefficient. General solutions \citep[e.\,g.,][]{len04:fdi,kwo05:tdi} have been derived for example in terms of the Fox H~function \citep{fox61:hfu}, which however is difficult to implement numerically \citep[e.\,g.,][]{ans12:fox}.

For the problem at hand, however, a classic solution equation is sufficient but equipped with time-dependent diffusion coefficients. In Eq.~\eqref{eq:frac}, this corresponds to the linear case, which is obtained for $\nu=1$ and $\mu=2$. The additional assumptions of homogeneity, $\vartheta=0$ so that $\kappa(x,t)=\kappa(t)$, and $\alpha=1$ yield the usual random walk.

If the only modification is the time-dependence in the diffusion coefficient in the form of a power-law, $\kappa=\xi t^\gamma$, a simple solution can be obtained as \citep[see][]{kwo05:tdi}
\be
f_{\text{1D}}(x,t)=\frac{1}{\sqrt{4\pi\xi t^{\gamma +1}}}\exp\!\left[-\frac{\left(1+\gamma\right)r^2}{4\xi\,t^{\gamma+1}}\right],
\ee
which has been derived before by \citet{bat52:di2} \citep[see also][]{hen84:dif}.

The generalization to three independent spatial dimensions can be easily achieved if the diffusion tensor is assumed to have a diagonal form, i.\,e., $\f\kappa=\text{diag}(\kappa_x,\kappa_y,\kappa\pa)$, which assumes the off-diagonal elements to vanish. These drift coefficients \citep[see, e.\,g.,][]{tau12:kpa} will be neglected, as drift motions are generally suppressed in turbulent magnetic fields \citep{min07:dri}. Their importance is, therefore, limited to cases where the mean magnetic field is curved, which results in drift velocities. Generally, the values for the drift coefficients are smaller than the parallel diffusion coefficient by orders of magnetiude.

Under these conditions, the solution to the three-dimensional diffusion equation,
\be
\pd[f_{\text{Diff}}]t=\left[\f\kappa(t)\cdot\nabla\right]\cdot\nabla f_{\text{Diff}}
\ee
decouples and is simply the product of the three one-dimensional solutions so that it reads
\be
f_{\text{Diff}}(\f r,t)=\prod_i\frac{1}{\sqrt{4\pi\xi_it^{\gamma_i+1}}}\exp\!\left[-\frac{\left(1+\gamma_i\right){r_i}^2}{4\xi_it^{\gamma_i+1}}\right]
\ee
for $i\in\{x,y,z\}$. The diffusion coefficients can be inserted individually as $\kappa_i(t)=\xi_it^{\gamma_i}$. It is this function that is used in text to fit the time profiles that are obtained in the \textsc{Padian} simulation code.

\section{Time-dependent diffusion coefficient: numerical approach}\label{appB}

From a numerical point of view, the diffusion equation has the special property that its spatial coordinate is unlimited and that the solution function has no boundary conditions except $f(x)\to0$ for $x\to\pm\infty$.

Therefore, a hyperbolic transformation in the spatial coordinate is done as $\zeta=\tanh(\eps x)$ with $\zeta\in[-1,1]$. This results in a distribution of the spatial grid points with a maximum density around $x=0$ that is exponentially decreasing toward $x\to\pm\infty$.

The diffusion equation can then be written as
\be
\pd[f]t=\eps^2\kappa(t)\left(1-\zeta^2\right)\pd\zeta\left[\left(1-\zeta^2\right)\pd[f]\zeta\right].
\ee
For the norm of the solution function, one has
\be
\uint\df x\;f(x,t)=\frac{1}{\eps}\int_{-1}^1\frac{\df\zeta}{1-\zeta^2}\;f(\zeta,t)\stackrel{!}{=}1,
\ee
which remains constant and therefore represents an additional constraint to the solution function. The remaining free parameter, \eps, was chosen so that (i) the largest $x$~values are 10~times larger than the detector position and (ii) the detector position falls within the domain where the tanh transformation results in a fine resolution.

To compute the numerical solution, the equation is discretized in time with an implicit Euler scheme with time step $\df t$. Consequently, in time $t_n$ with $f_n=f(\zeta,t_n)$ there holds
\be\label{eq:app:B:1}
f_n-\df t\left(\varepsilon^2 \kappa(t_n)\left(1-\zeta^2\right)\pd\zeta\left[\left(1-\zeta^2\right)\pd[f_n]\zeta\right]\right)=f_{n-1}.
\ee
The spatial discretization of this equation is done with the Ritz-Galerkin or finite element method \citep{Ritz1909,Galerkin1915}. The finite element method (FEM) solves instead of Eq.~\eqref{eq:app:B:1} the associated variational problem, which corresponds to the principle of virtual work, and reads for all test functions $v$ after integration by parts
\begin{align}
\int_{-1}^1\df\zeta\;f_n v&+\df t\int_{-1}^1\df\zeta\;\varepsilon^2 \kappa(t_n)\left(1-\zeta^2\right)^2\pd[f_n]\zeta\,\pd[v]\zeta\nonumber\\
&=\int_{-1}^1\df\zeta\;f_{n-1} v. \label{eq:app:B:2}
\end{align}

For the so-called $P_1$-FEM \citep{Brenner2008}, $f_n$ and $v$ are linear spline functions on $[-1,1]$. Therefore, the derivatives and integrals in \eqref{eq:app:B:2} can be easily computed leading in every time step $t_n$ to the linear system of equations
\be
\left(\mathsf M+\df t\,\mathsf S_n\right)\f f_n=\mathsf M\f f_{n-1},
\ee
where $\mathsf M$ is the so-called mass matrix, $\mathsf S$ the modified stiffness matrix due to the factor $\varepsilon^2 \kappa(t_n) (1-\zeta^2)^2$, and $\f f_n$ the coefficient vector for the linear spline function $f_n$.

Adding to this equation the constraint that the norm of the solution should be constant, the system is overdetermined, which can be solved with the method of ordinary least squares. There, the solution to an equation of the form $\mathsf A\f x=\f b$ is obtained by minimizing $\|\mathsf A\f x-\f b\|_2$ so that the solution is
\be
\f x=(\mathsf{A}{^\text T}\mathsf A)^{-1}\mathsf A^{\text T} \f b.
\ee
Because the time derivative was obtained as backward-Euler step, the solution is very robust with respect to the size of the time step, whereas it is crucial to apply the spatial transformation as described above.



\end{document}